\documentclass[a4paper]{jpconf}
\usepackage{graphicx}
\begin{document}
\title{Masses and magnetic moments of heavy flavour baryons in hyper central model}

\author{Bhavin Patel$^*$, Ajay Kumar Rai{$^\dag$$^\ddag$} and P C Vinodkumar$^*$}

\address{$^*$Department of Physics, Sardar Patel University,
Vallabh Vidyanagar, Anand-388 120, INDIA.\\
$^\dag$Applied Physics Department, Faculty of Technology \&
Engineering, M. S. University of Baroda, Vadodara - 390 001, Gujarat,INDIA.\\
$\ddag$ Present address Department of Applied Sciences and
Humanities, Sardar Vallabh National Institute of Technology,
Surat-395 007, Gujarat, INDIA}

\ead{azadpatel2003@yahoo.co.in}

\begin{abstract}
we employ the hyper central approach to study the masses and
magnetic moments of the baryons constituting single charm and
beauty quark. The confinement potential is assumed in the hyper
central co-ordinates of the coulomb plus power potential form.
\end{abstract}

\section{Introduction}
Recently, there is a renewed interest in the magnetic moments and
spectroscopy of heavy flavour baryons both experimentally and
theoretically \cite{Giannini2001,
  Ebert2005,Amand2006,Eduarev2006,Gunnar2001,Avery1995}. Many of the constituent quark models have provided the masses
of baryons and their magnetic moments correctly at the light
flavour sector. However many of these models do not provide the
form factors correctly that reproduces experimental data
\cite{Giannini2001} and for this reason alternate schemes to
describe the properties of baryons particularly in the heavy
flavour sector are being attempted \cite{Giannini2001}. It should
be mentioned that, hyper central potential contains the effects of
the three body force as it is suggested by lattice QCD
calculations \cite{Gunnar2001}. For the low-lying resonance states
it is good approximation to simply take the space wave functions
of the hyper Coulomb potential instead of the ones coming from the
numerical solution of the linear plus Coulomb potentials with
hyperfine interaction. We computed the masses of the charmed  and
beauty baryons under this scheme for different power indices
starting from 0.5 to 2.0.The magnetic moments of heavy flavour
baryons are computed based on the nonrelativistic quark model
using the spin-flavour wave functions of the constituting quarks
and their effective masses within the baryon.
\section{The Model}
The model Hamiltonian for the baryon is expressed in terms of the
Jacobin co-ordinates $(\rho,\lambda)$ as well as hyper central
co-ordinates (x) as be written as
\begin{equation}\label{eq:404}
H=\frac{P^2_\rho}{2\,m_\rho}+\frac{P^2_\lambda}{2\,m_\lambda}+V(\rho,\lambda)=\frac{P^2_x}{2\,m}+V(x)\\
\end{equation}\label{}
For the present study we considered the hyper central potential
as,  $ V(x)=-\frac{\tau}{x}+ \beta \cite{Ajay2005}.
x^\nu+\kappa+V_{spin}\left(x\right)$.The baryon masses are
computed as $M_B=\sum\limits_{i}m_i+\langle H \rangle=
\sum\limits_{i}m^{eff}_{i}$.The magnetic moment of baryons are
obtained in terms of its constituent quarks as $
\mu_B=\sum\limits_{i}\left<\phi_{sf}\mid\mu_{i}
\vec{\sigma}_{i}\mid\phi_{sf}\right>$ where $
\mu_{i}=\frac{e_{i}}{2m_{i}^{eff}}$.  Here $m_i$, $e_{i}$ and
$\sigma_{i}$ represent the mass, charge and the spin of the quark
constituting the baryonic state and $\mid\phi_{sf}>$ represents
the spin-flavour wave function of the respective baryonic state.
Our model parameters are $m_c= 1394 MeV$,$m_b= 4510 MeV$,$m_u= 338
MeV$,$b= 13.6$,$\frac{\beta}{m \tau}=1 (MeV)^\nu$, $A=140.7 MeV$
and $\alpha=850 MeV$.

\begin{table*}
\begin{center}
\caption{Masses of charm and beauty  baryons
 in MeV}
 \label{tab:01}
%\begin{indented}
%\item[]
\begin{tabular}{lccccccccc} \hline \hline
&\multicolumn{5}{c}{\textbf{\underline{\, \,\,\,\, \,\,\,\, \,\,\,\, \,Potential index $\nu$\, \, \,\,\,\, \,}}}&&&\\
Baryon &0.5&0.7&1.0&1.5&2.0&\cite{Bowler1998}&\cite{PDG2006}&\cite{Gorelov2007}\\
\hline
$\Sigma^{++}_{c}$&2550&2473&2443&2436&2436&2460$\pm$80&2454$\pm$0.18&$-$\\
$\Sigma^{*++}_{c}$   &2618&2538&2506&2499&2498&2440$\pm$70&2518.4$\pm$0.6&$-$\\
\hline
$\Sigma^{+}_{b}$&5871&5812&5787&5780&5780&5770$\pm$70&$-$&$5808^{+02}_{-2.3}\pm$1.7\\
$\Sigma^{*+}_{b}$   &5808&5837&5810&5802&5802&5780$\pm$70&$-$&$5829^{+1.6}_{-1.8}\pm$1.7\\
\hline

\end{tabular}
\end{center}
\end{table*}

\begin{table*}
\begin{center}
\caption{Magnetic moments of charm and beauty baryons
 in terms of Nuclear magneton $\mu_{N}$ }
 \label{tab:02}
%\begin{indented}
%\item[]
\begin{tabular}{lcccccccc} \hline \hline
&\multicolumn{5}{c}{\textbf{\underline{\, \,\,\,\, \,\,\,\, \,\,\,\, \,\,\,\, \,\,\,Potential index $\nu$\, \, \,\,\,\, \,\, \, \,}}}&&&\\
Baryon &0.5&0.7&1.0&1.5&2.0&RQM\cite{Amand2006}&NRQM\cite{Amand2006}\\
\hline
$\Sigma^{++}_{c}$&1.8809&1.9394&1.9635&1.9688&1.9692&1.760&1.860\\
$\Sigma^{*++}_{c}$   &3.2806&3.3837&3.4272&3.4373&3.4379&$-$&$-$\\
\hline
$\Sigma^{+}_{b}$&2.1995&2.2216&2.2314&2.2339&2.2341&2.070&2.010\\
$\Sigma^{*+}_{b}$   &3.2806&3.3837&3.4272&3.4373&3.4379&$-$&$-$\\
\hline
\end{tabular}
\ \ .\\ $^*$indicates $J^P=\frac{3}{2}^+$ state.
\end{center}
\end{table*}
\section{Result and Discussion}
The computed masses and magnetic moments of the J$=\frac{1}{2}^+$
and J$=\frac{3}{2}^+$ states of the single charm and beauty
flavour baryons are listed in Table \ref{tab:01} and \ref{tab:02}.
The masses and magnetic moments of the single heavy flavour
baryons are found to be in accordance with other
 model predictions. It is important to see that the baryon mass do not
change appreciably beyond the potential power index $\nu
> 1.0$ (See Table \ref{tab:01}) and the magnetic
moment predicted in our model do not vary appreciably with
different choices of $\nu$ running from 0.5 to 2.0 (See Table
\ref{tab:02}).
\section*{Acknowledgement} One of the author P. C. Vinodkumar
acknowledge the financial support from UGC, Government of India
under a Major research project F. 32-31/2006 (SR).

\smallskip

\end{document}